\documentstyle [12pt]{article}
\newcommand{\cs}[3]{{{#3} \brace {#1 #2}}}

\newcommand{\edf}{\ {\mathop{=}\limits^{\rm def}}\ }

\newcommand{\al}{\alpha}

\begin{document}

\begin{center}
{\bf {DARK ENERGY: A MISSING PHYSICAL INGREDIENT}}
\end{center}
\begin{center}
{  M. I. Wanas \\  {\it Astronomy Department, Faculty of
Science, Cairo University, Giza, Egypt.}\\
{\it and The Egyptian Relativity Group}\\
 e-mail: wanas@frcu.eun.eg\\
http://www.erg.eg.net }
\end{center}
 {\bf{Abstract}}\\
Recent observation of supernovae  type Ia show clearly that there is
a large scale repulsive force in the Universe. Neither of the four
known fundamental interactions can account for this repulsive force.
  Gravity is
known to be the interaction responsible for the large scale
structure and evolution of the Universe. The problem with gravity is
that it gives rise to a force which is attractive only. Gravity
theories, including General Relativity, deals with gravity as an
attractive force. Although this is consistent with our experience in
the solar system and other similar astrophysical systems, gravity
theories fail to account for SN type Ia observation. So, we are in a
real problem concerning the interpretation of these observation.
This problem is only ten years old. In order to go out of this
problematic situation, scientists have suggested the existence of a
type of energy in the Universe that is responsible for the above
mentioned repulsive force. They have given this type of energy  the
exotic term {\it "Dark Energy"}. Although this type of energy forms
more than two thirds of the energetic contents of our Universe, its
reasonable nature is missing in  all gravity theories.

The aim of the present work is to review the present status of the
problem of dark energy. Also, to suggest a new geometric solution
for this problem.

\section{Introduction}
In the last ten years, astronomical observations show that some
phenomena are not consistent with the present accepted physical
theories. Among these observations: \\
1. The supernovae (SN) type Ia observations [1],which indicate very
clearly that our Universe is in an accelerating expansion phase. The
problem is that this type of expansion necessities the presence of a
large scale repulsive force in the Universe. Assuming that gravity
plays the main role in the structure and evolution of the Universe,
one cannot interpret the above mentioned observations. The reason is
that gravity, as known in the solar system, gives rise to an
attractive force only, which is treated as a physical {\it fact}.
Gravity theories, including general relativity (GR), are constructed
taking this {\it fact} into account. More precisely, the dynamical
equations of FRW-models (with vanishing cosmological constant) give
rise to a de-accelerating Universe. This is a real discrepancy
between theory and
observations. \\
2. The rotation curves of spiral galaxies [2], which concern
observations of star rotational velocities in such type of galaxies.
The curves resulting from observational (giving the relation between
the distance of the star from the center of galaxy and its rotation
velocity), show a flat behavior for the outer regions of the galaxy,
while the curves resulting from theories, including the orthodox GR
, are bent towards the x-axis (the axis giving the distance of the
star from the center  of the galaxy). This gives another discrepancy
between observations and known
physics.\\
3. The observation of the velocities of spacecrafts, Pioneer 10, 11
[3]. These two spacecrafts were launched in March 1972 and April
1973, respectively. Both vehicles have an apparatus emitting radio
waves with certain wave length. The comparison between  radio waves
received from these vehicles and radio waves from an identical
apparatus, on the Earth's surface, gives the radial velocities of
each vehicle, via the red-shift phenomena, for the two vehicles, are
found to be different from those obtained from gravity theories.
This represents a third discrepancy between experiment and gravity
theories.

It seems that the above mentioned problems are closely connected to
each other. All are on scales larger than the solar system scale and
have some relation to gravity theories.  These problems indicate
clearly that an important ingredient is missing from gravity
theories. The correct identification of this ingredient,  leading to
a solution of one of the above mentioned problems, would
automatically lead to solutions of the other problems, if they are
connected.

The aim of the present work is to throw some light on possible
ingredients,  missing from gravity theories, concentrating on the
first problems. In section 2 we give  brief survey of different
interpretations of dark energy. In section 3 we introduce a
principle leading our investigations . In section 4. we give a brief
review on the  geometry appropriate for the suggested solution. The
paper is discussed and concluded in section 5.

\section{Different Interpretations of Dark Energy}
Since the discovery of the accelerating expansion of the Universe,
many authors have suggested different solutions for this problem.
The solutions suggested imply the existence of some type of energy
with negative pressure. This type of energy is known in the
literature as {\it "Dark Energy"}. We are going to review briefly
different solutions, of this problem, proposed to probe the nature
of such peculiar type of energy.

The field equations of GR can be written as,  $$ R_{\mu \nu}-
\frac{1}{2}g_{\mu \nu}R = -\kappa T_{\mu \nu},  \eqno{(1)}$$ where
$R_{\mu \nu}$ is Ricci tensor, $R$ is Ricci scalar and $T_{\mu \nu}$
is the material energy tensor. In case of FRW-models with the metric
($x^{1}\equiv r, x^{2}\equiv \theta, x^{3}\equiv \phi \ \ \& \ \
x^{4}\equiv t $),  $$ ds^{2}= dt^{2} - \frac{a^{2}(t)}{1+ \frac{k
r^{2}}{4}} (dr^{2}+ r^{2}d^{2}\theta + r^{2}sin^{2} \theta
d^{2}\phi^{2}), \eqno{(2)} $$ where $a(t)$ is the scale factor and
$k$ is the curvature constant, the field equations (1) give rise to
the dynamical equations,   $$ \frac{\dot{a}^{2}}{a^{2}} =
\frac{8}{3} \pi \rho_{o}- \frac{k}{a^{2}},  \eqno{(3)}$$
$$ \frac{\ddot{a}}{a} = - \frac{4 }{3} \pi (\rho_{o}+3p_{o} ).\eqno{(4)} $$

It is clear from (4) that , for the proper density $\rho_{o}>0$ and
for the proper pressure $p_{o}\geq0$, $\ddot{a}$ is negative
(de-acceleration), which contradicts SN type Ia observations, as
mentioned above. This problem may be solved if there is an
additional term on the right hand side of (4) compensating the
gravitational attraction. This can be achieved using one of the
following suggestions:\\ \\
1. By inserting a term, {\it{the
cosmological term}} ($g_{\mu \nu}\Lambda$), into the definition of
the Einstein tensor as done by Einstein himself (cf. [4]). In this
case, GR field equations will take the form
$$ R_{\mu \nu}-
\frac{1}{2}g_{\mu \nu}(R-2\Lambda) = -\kappa T_{\mu \nu}. \eqno{(5)}
$$ This will add a term in the dynamical equations compensating
gravitational attraction to give, qualitatively, the observed
phenomena. \\ \\
2. By imposing an equation of state, with negative pressure, of the
form, $$ p_{o} =\omega \rho_{o},  \eqno{(6)}$$ where $\omega$ is a
negative parameter. It is to be considered that the use of (5),  to
construct world models, corresponds to $\omega=-1$. The cases with
$\omega<-1$ are known as {\it{"phantom energy"}} (cf. [5]), while
the cases with $\omega>-1$ are known as {\it{"quintessence energy"}}
(cf. [6]). Belongs to this class of suggestions the use of an
equation of state of the form,  $$ p_{o} =-\frac{A}{
\rho_{o}^{\alpha}}, \eqno{(7)}$$ where $A(>0)$ and $\alpha (\geq0)$
are parameters. This case is known in the literature as {\it
{"Chaplygin gas"}} (cf. [7]). \\ \\
3. "{\it{Modified Gravity}}" (cf.
[8]): It is well known that in constructing the field equations of
GR (4), using an action
 principle, the Lagrangian function contains a term linear in Ricci
 scalar,  $R$. In the modified gravity theories, it is suggested that this term
 is replaced by any function $f(R)$ including quadratic and negative powers of
 $R$.\\ \\
 4. In this class of suggestions it is claimed that the introduction
 of {\it{extra dimension}} would solve the problem (cf. [9]). The
 resulting theory is of Kaluza-Klein type.

\section{The Interaction Principle }
In the previous section, a summary of different suggestions to
understand the nature of dark energy are given. Some suggestions
retain GR written in Riemannian geometry, while others modify GR, in
Riemannian geometry also. The common feature in all attempts,
summarized above,  are carried out in the context of Riemannian
geometry. The main assumption in this case is that:\\
\\{\underline{"Riemannian space gives a complete representation of
the physical world }}\\ {\underline{including space and time"}}.\\
\\
However, we are going to relax this assumption in the present work.

Furthermore, in the present attempt, in order to explore what is
missing behind the exotic term "{\it{dark energy}}" we are going to
use the {\it interaction principle} [10]. This can be summarized as
follows: \\ \\{\it{"Physical phenomena are just interactions between
space-time properties and the intrinsic properties of the elementary
constituents of matter (energy)"}}. \\ \\ Now assuming the validity
of this principle and that the ingredient behind dark energy is an
interaction of the type mentioned in this principle, one way to
explore this interaction is to relax the assumption underlined
above, in order to extend the geometric structure. This is because
Riemannian space is limited and has no sufficient structure to apply
the interaction principle satisfactorily. Its linear connection is
symmetric which implies the vanishing of the torsion of space-time.
The introduction of  a non-symmetric linear connection will provide
us with a geometric structure with a non-vanishing torsion. This
will produce a wider space, more thorough than the Riemannian one,
especially when it possesses a non-vanishing curvature. The extra
structure of this space, can give more degrees of freedom which
facilitates the application of the interaction principle. This would
throw some light on the nature of dark energy. The solutions
suggested in this case would be a pure geometric one.
\section{An Appropriate Geometric Structure}
In order to explore what is the missing ingredient behind the
peculiar assumption of {\it dark energy} we are going to apply
"{\it{the interaction principle}}". For this reason, it is
preferable to use a complete geometric structure with the following
general properties:\\ \\
1. It should have a metric tensor since conventional gravity is
proved to be a metric phenomena. \\ \\
2. Its linear connection should be non-symmetric in order to have a
non-vanishing torsion. This will extend the geometric structure. \\
3. It should have a non-vanishing curvature, as well. \\ \\
These properties characterize a wide class of geometric structures
known in the literature as the "Riemannian-Cartan geometry". We are
going to use a certain structure of this class, known as the
"{\it{Parameterized Absolute Parallelism (PAP)-geometry}}" [11].
Calculations within the context of this structure are easy and the
physical meaning are clear. The non-symmetric linear connection of
this structure can be written as,
 $$\hat\Gamma^{\alpha}_{\mu \nu}=
\cs{\mu}{\nu}{\alpha}+ b \gamma^{\alpha}_{. \mu \nu}, \eqno{(8)}$$
where $\cs{\mu}{\nu}{\alpha}$ is the Christoffel symbol of the
second kind defined in terms of the metric of the space, $b$ is a
parameter and $\gamma^{\alpha}_{. \mu \nu}$ is the contortion of
AP-structure. The torsion of the  PAP-structure can be defined by
(AP-quantities are without hats)$$\hat \Lambda^{\alpha}_{. \mu \nu}
\edf \hat\Gamma^{\alpha}_{. \mu \nu}-\hat\Gamma^{\alpha}_{. \nu
\mu}= b (\gamma^{\alpha}_{. \mu \nu}- \gamma^{\alpha}_{. \nu \mu} ),
$$ i.e.
$$\hat\Lambda^{\alpha}_{. \mu \nu} = b \Lambda^{\alpha}_{. \mu \nu} . \eqno{(9)}$$
The Curvature corresponding to the non-symmetric connection (8) is
in general non-vanishing and given by,
$$\hat B^{\alpha}_{. \mu \nu \sigma}=\hat\Gamma^{\alpha}_{.\mu \sigma, \nu}- \hat\Gamma^{\alpha}_{.\mu \nu ,\sigma}
 +\hat \Gamma^{\epsilon}_{. \mu\sigma} \hat\Gamma^{\alpha}_{. \epsilon \nu}-
 \hat \Gamma^{\epsilon}_{. \mu\nu} \hat\Gamma^{\alpha}_{. \epsilon
 \sigma},
 \eqno{(10)}$$ which can be written in the form [12].
 $$\hat B^{\alpha}_{. \mu \nu \sigma}=  R^{\alpha}_{. \mu \nu \sigma}+ b \hat Q^{\alpha}_{. \mu \nu \sigma} , \eqno{(11)} $$
where $$R^{\alpha}_{. \mu \nu \sigma} =
\cs{\mu}{\sigma}{\alpha}_{,\nu}-\cs{\mu}{\nu}{\alpha}_{,\sigma}+
\cs{\mu}{\nu}{\epsilon}\cs{\epsilon}{\sigma}{\alpha}-
\cs{\mu}{\sigma}{\epsilon}\cs{\epsilon}{\nu}{\alpha} \eqno{(12)}
$$ and $$\hat Q^\alpha_{.\ \mu \sigma \nu} \edf {\gamma}^{\stackrel{\al}{+}}_{{.\
{\stackrel{\mu}{+}}{\stackrel{\nu}{+}}} | \sigma} -
{\gamma}^{\stackrel{\alpha}{+}}_{.\
{{\stackrel{\mu}{+}}{\stackrel{\sigma}{-}}} | \nu} + b~(
{\gamma}^{\beta}_{. \mu \sigma} {\ } {\gamma}^{\alpha}_{. \beta \nu}
{\ } - {\gamma}^{\beta}_{. \mu \nu} {\ } {\gamma}^{\alpha}_{. \beta
\sigma}). \eqno{(13)} $$ The general path equation for this
structure can be written as [13],$$ {\frac{dZ^\mu}{d\tau}} +
\cs{\alpha}{ \beta}{\mu} Z^\alpha Z^\beta = - b~~ \Lambda^{~ ~ ~ ~
\mu}_{(\alpha \beta) .} ~~Z^\alpha Z^\beta, \eqno{(14)}$$ where
$\tau$ is a parameter characterizing the path and $Z^{\mu}$ is its
tangent.\\

{\bf{Comments on the PAP-structure}}: \\  1. It represents a class
of Riemann-Cartan geometry. It has simultaneously non-vanishing
parameterized torsion (9) and
curvature(10). \\ \\
2. Although the curvature (10) is non-linear in the connection (8),
it can be slitted into two parts: the first is a pure function of
Christoffel symbol only,  $R(\{\})$, which is the
Riemann-Christoffel curvature tensor; while the second is a fourth
order tensor, function of the contortion (or the torsion) only,
$\hat Q(\gamma)$ . This is an important property in physical
applications which is
discussed later. \\ \\
3(a)- For $b=0$ the PAP-structure becomes Riemannian. Equations
(9),(11)and (14) will reduce respectively, to
$$\hat\Lambda^{\alpha}_{. \mu \nu}= \Lambda^{\alpha}_{. \mu \nu}=0 , \eqno{(15)}$$
$$\hat B^{\alpha}_{. \mu \nu \sigma}=  R^{\alpha}_{. \mu \nu \sigma} ,\eqno{(16)} $$
$$ {\frac{dZ^\mu}{d\tau}} +
\cs{\alpha}{\beta}{\mu} Z^\alpha Z^\beta = 0. \eqno{(17)}$$ So, any
field theory constructed in the PAP-space can be easily compared
with orthodox GR upon taking to $b=0$. Furthermore, equation (14) is
reduced to the ordinary geodesic equation (17), under the same
condition. This equation gives rise to the attractive force of
conventional gravity. \\ \\
3(b)- For $b=1$ the PAP-structure reduces to the conventional
AP-structure (cf.[14]) with a non-vanishing torsion
$$\hat
\Lambda^{\alpha}_{. \mu \nu} \Rightarrow \Lambda^{\alpha}_{. \mu
\nu} = \Gamma^{\alpha}_{. \mu \nu}- \Gamma^{\alpha}_{. \nu \mu}=
(\gamma^{\alpha}_{. \mu \nu}- \gamma^{\alpha}_{. \nu \mu} ),
\eqno{(18)} $$ but with a vanishing curvature [14].
$$\hat B^{\alpha}_{. \mu \nu \sigma}\Rightarrow B^{\alpha}_{. \mu \nu \sigma}
=R^{\alpha}_{. \mu \nu \sigma}+  Q^{\alpha}_{. \mu \nu
\sigma}\equiv0. \eqno{(19)}$$ The last relation is very important.
Although the curvature $B(\Gamma)$, given by (19), is an identically
vanishing tensor, giving rise to a flat space, its constituents
$R(\{\})$ and $Q(\gamma)$ are not vanishing objects. The tensor
$Q(\Gamma)$, defined by (13) with $b=1$, compensates the curvature
$R(\{\})$ of the space-time in such a way that $B(\Gamma)$ vanishes,
as shown by (19). For this reason the tensor $Q(\gamma)$ is called
{\underline{\it the anti-curvature}} or {\underline{\it the additive
inverse}} of the curvature tensors. Both tensors satisfy Bianchi
differential identity ,  whose contracted form can be written, for
these tensors, as [15]
$$ (R^{\mu \nu}- \frac{1}{2}g^{\mu \nu}R)_{;\mu} \equiv 0, \eqno{(20)}$$

$$ (Q^{\mu \nu}- \frac{1}{2}g^{\mu \nu}Q)_{;\mu} \equiv 0 .\eqno{(21)}$$
The identity (20) is interpreted, in the context of geometrization
scheme, as a geometric representative of some conservation law, so
is (21). For this reason we call the quantities between brackets in
equations (20) and (21) the {\it{curvature energy}} and the
{\it{torsion energy}}, respectively [15]. Torsion energy is a type
of energy that is mainly caused by torsion, since the {\it
anti-curvature} tensor $Q(\gamma)$ is purely made of the contortion
(or the torsion). It has been shown in [10] that torsion energy is
associated with a repulsive force giving rise to
{\it{anti-gravity}}. \\ \\
4. The term on the right hand side of the PAP-path equations (14)
has been interpreted physically as representing a type of
interaction between the torsion of the background space-time and the
quantum spin of the moving particle [13]. The linearized form of
this equation gives rise to the generalized potential,
$$\Phi=\Phi_{N}+ \Phi_{T}, \eqno{(22)}$$ where $\Phi_{N}$ is the
Newtonian gravitational potential and
$$\Phi_{T}=- b \Phi_{N}, \eqno{(23)}$$  is the torsion anti-gradational
potential. It is clear that, as $\Phi_{N}$ gives rise to an
attractive force, $\Phi_{T}$ will give rise to a repulsive one.

\section{Conclusion}
1. The existence of torsion, in the space-time structure, gives rise
to a repulsive force. Consequently torsion energy is an energy
connected to this force, which gives rise to a negative pressure. \\
\\2. The missing physical ingredient is the {\it spin-torsion} interaction
and
consequently dark energy is nothing more than torsion energy. \\ \\
3. Since electromagnetic phenomena are the results of interaction
between the electromagnetic field and the electric charge (an
elementary particle intrinsic property), then one can conclude,
using the {\it interaction principle}, that the electromagnetic
field is a space-time property. Some authors have succeeded in
representing the electromagnetic field as a geometric property
(cf.[16], [17]) with successful applications (cf.
[18], [19], [20]).\\ \\
4. The main assumption, given in section 3, should now be replaced
by:
\\{\underline{"A space with simultaneously non-vanishing curvature and torsion gives  }}\\
 {\underline{a complete representation of
the physical world including space and time"}}.

\section*{Acknowledgements}
The author would like to thank the Organizing Committee of NUPPAC'07
and Professor Dr. M.N.H. Comsan for inviting him to give this talk.
\section*{References}
{[1] J.L. Tonry et al. (2003) Ap. J.
{\bf 594} 1.}\\
{[2] A.A. Kirilov  and D. Turaev,  (2006) Mon. Not. Roy. Astron.
Soc. {\bf 371}

L31.}\\
{[3] M.M. Neito  and J.D. Anderson  (2007) arXiv:0709.3866.}\\
{[4] S.M. Carroll, {\it "The Cosmological Constant"},
http://livingreviews.org/

Irr-2001-1.}\\
{[5] J. Cepa (2004) Astron. Astrophys. {\bf 422} 831.}\\
{[6] L.P. Chimento, A.S. Jakubi and D. Pavon(2000) arXiv:astro-ph/0010079.}\\
{[7] A. Dev, J.S. Alcaniz and D. Jain (2003) Phys. Rev. {\bf D67} 023515.}\\
{[8] A. Lue, R. Scoccimarro and G. Starkman (2004) Phys. Rev. {\bf D69} 044005.}\\
{[9] D. Ponigrahi, Y.Z. zhang and s. Chatterjee (2006) Int. J. Mod. Phys.

{\bf A21} 6491.}\\
{[10] M.I.Wanas (2007) Int. J. Mod. Phys. {\bf A22} 5709; arXiv:0802.4104.}\\
{[11] M.I.Wanas (2000) Turk. J. Phys. {\bf 24} 473; gr-qc/0010099.}\\
{[12] M.I.Wanas (2003) Proc. XXV Int. Workshop {\it "Fundamental
Problems of

High Energy Physics and Field Theory"}, p. 315; gr-qc/0506081.}\\
{[13] M.I.Wanas (1998) Astrophys. Space Sci. {\bf 258} 237; gr-qc/9904019.}\\
{[14] M.I.Wanas (2001) Cercet. Stiin. Ser.Mat. {\bf 10} 297; gr-qc/0209050.}\\
{[15] M.I.Wanas (2007) arXiv:0705.2255.}\\
{[16] G.I. Shipov (1998) {\it "The Theory of Physical Vacuum"}, English ed. Moscow.}\\
{[17] F.I. Mikhail and M.I. Wanas (1977) Proc. Roy. Soc. Lond.
{\bf A356} 471.}\\
{[18] F.I. Mikhail and M.I. Wanas (1981) Int. J. Theoret. Phys.
{\bf 20} 671.}\\
{[19] M.I. Wanas (1985) Int. J. Theoret. Phys.
{\bf 24} 639.}\\
{[20] M.I. Wanas (2007) Int. J. Geom. Meth. Mod. Phys.
{\bf 4} 372;

gr-qc/0703036.}\\
\end{document}